\documentclass[aps,prl,twocolumn,showpacs,showkeys,amssymb,amsmath,superscriptaddress]{revtex4-1}
\usepackage{graphicx}

\newcommand{\V}[1]{\mathbf{#1}} 
 
\newcommand\Alfven{Alfv\'en }
\newcommand\Alfvenic{Alfv\'enic }
\newcommand{\figref}[1]{Fig.~\ref{#1}}

\newcommand{\XCC}{\mbox{$C(\delta n, \delta B_{\parallel})~ $}} 
\newcommand{\dnr}{\mbox{$\langle \delta n \rangle_{r}$}} 
\newcommand{\dbr}{\mbox{$\langle \delta B_{\parallel} \rangle_{r}$}} 
\usepackage{color}

\begin{document}

\title{The slow-mode nature of compressible wave power in solar wind turbulence}

\author{G.~G. Howes}
\affiliation{Department of Physics and Astronomy, University of Iowa, Iowa City, Iowa 52242, USA.}
\author{S.~D.~Bale}
\affiliation{Space Sciences Laboratory, University of California, Berkeley, 
California 94720-7450, USA.}
\affiliation{Department  of Physics, University of California, Berkeley, California, 94720-7300, USA.}
\author{K.~G. Klein}
\affiliation{Department of Physics and Astronomy, University of Iowa, Iowa City, Iowa 52242, USA.}
\author{C.~H.~K.~Chen}
\affiliation{Space Sciences Laboratory, University of California, Berkeley, 
California 94720-7450, USA.}
\author{C.~S.~Salem}
\affiliation{Space Sciences Laboratory, University of California, Berkeley, 
California 94720-7450, USA.}
\author{J.~M. TenBarge}
\affiliation{Department of Physics and Astronomy, University of Iowa, Iowa City, Iowa 52242, USA.}

\date{\today}

\begin{abstract}
We use a large, statistical set of measurements from the \emph{Wind}
spacecraft at 1~AU, and supporting synthetic spacecraft data based on
kinetic plasma theory, to show that the compressible component of
inertial range solar wind turbulence is primarily in the kinetic {\em
slow} mode.  The zero-lag cross correlation \XCC between proton
density fluctuations $\delta n$ and the field-aligned (compressible)
component of the magnetic field $\delta B_{\parallel}$ is negative and
close to $-1$.  The typical dependence of $C(\delta n,\delta
B_{\parallel})$ on the ion plasma beta $\beta_i$ is consistent with a
spectrum of compressible wave energy that is almost entirely in the
kinetic slow mode.  This has important implications for both the
nature of the density fluctuation spectrum and for the cascade of
kinetic turbulence to short wavelengths, favoring evolution to the
kinetic Alfv\'en wave mode rather than the (fast) whistler mode.
\end{abstract}

\pacs{96.50.Ci, 94.05.Lk, 52.35.Ra}

\maketitle 

\emph{Introduction.}---
The inertial range of solar wind turbulence is comprised of a mixture
of incompressible and compressible motions, with at least 90\% of the
energy due to the incompressible component \cite{Bruno:2005}.  If
these fluctuations are interpreted as some mixture of the three MHD
linear wave modes, then \Alfven waves are the dominant
incompressible component, while slow and fast MHD waves make up the
compressible component. These modes are distinguished by the
correlation between the density and parallel magnetic field
fluctuations: fast waves are positively correlated, slow waves are
negatively correlated, and the density and parallel magnetic field
fluctuations are both zero for
\Alfven waves. As the wave amplitude is increased to nonlinear levels,
even to the limit that they form discontinuities or shocks, these
qualitative properties persist, corresponding to tangential and
rotational discontinuities or fast and slow shocks
\cite{Baumjohann:1996}.  The MHD limit of strong collisionality, however, is not
valid in the solar wind; therefore, collisionless kinetic theory is
necessary to determine the properties of the wave modes. Each of the
kinetic versions of the MHD linear wave modes, determined using the
Vlasov-Maxwell linear dispersion relation, retain the qualitative
correlations between the density and parallel magnetic field
fluctuation described above \cite{Klein:2011}. In addition, these
kinetic counterparts to the compressible modes may suffer damping from
collisionless mechanisms
\cite{Barnes:1966}.

Compressible fluctuations at inertial range scales $\lambda$ in the solar wind
($10^{-4}$ Hz $\lesssim f_{sc} \sim v_{sw}/\lambda \lesssim 1$ Hz; $v_{sw}$ is
the solar wind speed, $f_{sc}$ is the Doppler-shifted frequency in the spacecraft
frame) have been studied
extensively, often interpreted as a mix of magnetoacoustic
(fast MHD) waves and pressure-balanced structures
(PBSs) \cite{Tu:1995,Bruno:2005}.  A PBS was first observed as an
anti-correlation of thermal pressure and magnetic pressure
at timescales of 1~h \cite{Burlaga:1970}, and subsequent
investigations found a similar anti-correlation between the density
and magnetic field magnitude
\cite{Vellante:1987,Roberts:1990}. Theoretical studies of compressible 
MHD fluctuations in the low-Mach number, high-$\beta$ limit
interpreted these anti-correlated density-magnetic field strength
observations as nonpropagating ``pseudosound'' density fluctuations
\cite{Matthaeus:1991}. A more comprehensive investigation 
confirmed the general density-magnetic field anti-correlation, but
also identified a few positively correlated intervals consistent
with the magnetosonic (fast MHD) wave \cite{Tu:1994}. Analysis of
\emph{Ulysses} observations found evidence for PBSs at inertial range
scales in the high latitude solar wind
\cite{McComas:1995,Reisenfeld:1999,Bavassano:2004}. Studies of the
electron density up to $f_{sc}=2.5$~Hz also found pressure balanced
structures, but interpreted these as ion acoustic (slow MHD) waves,
and recognized that PBSs are simply the ion acoustic (slow MHD) wave
in the perpendicular wavevector limit \cite{Kellogg:2005}. Recently,
measurements of the anti-correlation between electron density and
magnetic field strength indicated the existence of PBSs over
timescales ranging from 10$^3$~s down to 10~s \cite{Yao:2011}.

This Letter demonstrates that the compressible fluctuations in the
inertial range are almost entirely kinetic slow wave fluctuations,
suggesting that little turbulent energy is transferred from large
scales to whistler fluctuations below the ion gyroscale.  First, we
show that the density-magnetic field cross-correlation \XCC in the
solar wind is  $\simeq -1 $ and increases slightly with ion
plasma beta, $\beta_i$.  Then we demonstrate excellent agreement with
synthetic (eigenfunction) data in which less than 10\% of the
compressible energy is due to fast waves.

\begin{figure}
\resizebox{3.3in}{!}{\includegraphics{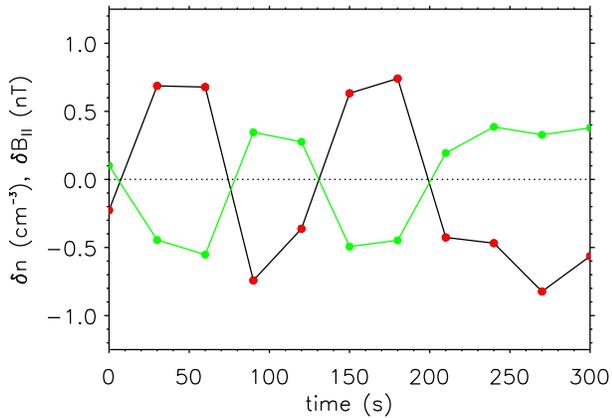}}
\caption{ \label{fig:waveform} Example of waveforms of proton density 
fluctuations $\delta n$ (black with red dots) and parallel magnetic field $\delta
B_\parallel$ (green) fluctuations.  This interval has \dnr $\sim$ 0.6
cm$^{-3}$,  \dbr $\sim$ 0.4 nT, and  \XCC $\sim$ -0.97, 
however, even intervals with smaller density fluctuations
exhibit significant anti-correlations.}
\end{figure}

\emph{Measurements.}--- We use measurements from the Magnetic Field 
Investigation (MFI) \cite{Lepping:1995p6447} and the Three Dimensional
Plasma (3DP) experiment \cite{Lin:1995p934} on the \emph{Wind}
spacecraft, in the unperturbed solar wind at 1 AU, during the years
1994-2004.  The magnetic field is sampled at either 11 or 22 vectors/s
(depending on the spacecraft telemetry rate) then averaged down to the
spacecraft spin period (3~s).  The ion moments are computed on board
the spacecraft at 3~s cadence; protons are separated from alpha
particles by a fixed energy interval (which is occasionally adjusted
in flight software) during the moment calculation; the solar wind
alpha particle abundance is typically 3-5\% of the proton number
density.  We select 1,089,491 300-s intervals of ambient solar wind
data (corresponding to spatial intervals of approximately $L \approx$
450 km/s $\times$ 300 s = 135,000 km $\sim$ 1350 $\rho_{i}$, where
$\rho_i$ is the ion Larmor radius) and the data is decimated by a
factor of 10 (to 30~s cadence).  Therefore, our data correspond to
inertial range scales of approximately $k \rho_i \in (5 \times
10^{-3}, 5 \times 10^{-2})$.  After selection, the magnetic field data
are averaged in 100-s windows to compute the local mean field
$\V{B}_0$.  The fluctuation field $\delta \V{B}$ is created by
subtracting $\V{B}_0$ and  then rotated to a field-aligned
coordinate (FAC) system defined by the $\V{B}_0$ direction.  In this
new system, there is a compressible field fluctuation $\delta
B_{\parallel}$ and shear components $\delta B_{\perp,1}$ and $\delta
B_{\perp, 2}$.  The most probable amplitude of the shear component $\delta B_{\perp}
= (\delta B^2_{\perp,1} + \delta B^2_{\perp,2})^{1/2}$ is approximately
3.4 times greater than most probable $\delta B_{\parallel}$ and corresponds to
the Alfv\'enic component of the turbulence.

Proton density data $\delta n = n - n_0$ is detrended over the same
time intervals.  Proton density is an integral over the (3 s)
distribution function $f(\V{v})$, computed from particle flux
measurements in 16 individual energy channels
\cite{Lin:1995p934}.  Since the counts are digitized discretely and
the energy channels are also discrete (with $\Delta E/E \sim 0.2$),
the 3DP proton density moments have a finite dynamic range.  To assess
this, we evaluated the joint probability distribution
(not shown) of \dnr ~and \dbr, the RMS values.  Below values of \dnr $
\approx$ 0.5 cm$^{-3}$, the joint pdf reverts from being
well-correlated to a broader set of values.  Furthermore, the
histogram of \dnr~alone shows clearly an artificial (non-Poisson)
lower cutoff at around this value, a cutoff not seen in \dbr.  We take \dnr $
\approx$ 0.5~cm$^{-3}$ as the ``noise'' level of the density
fluctuation measurement and restrict our analysis to intervals in
which \dnr $ \ge $ 0.5~cm$^{-3}$.  This restriction reduces the
dataset from 1,089,491 to 119,512 data intervals and biases our sample
to higher absolute densities.  The most probable values of absolute density are 
$\sim$4 cm$^{-3}$ for the full and $\sim$12 cm$^{-3}$ for the 
thresholded dataset. 
The distribution of plasma $\beta_i$ is unaffected by the thresholding. 

\begin{figure}
\resizebox{3.4in}{!}{\includegraphics{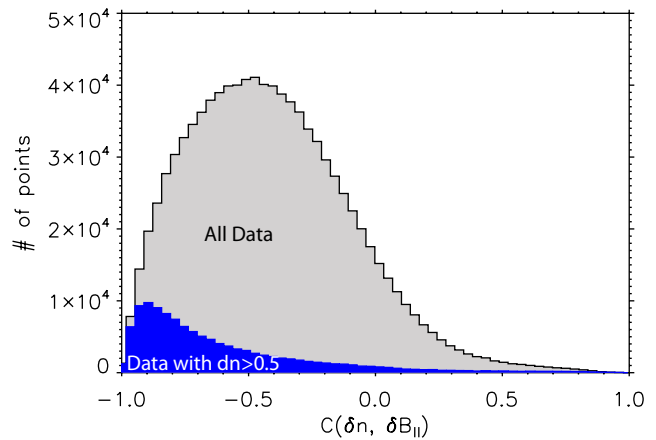}}
\caption{ \label{fig:histo} Histogram of the   
cross-correlations $C(\delta n, \delta B_{\parallel})$ (grey, 1,089,491
points total) and those above the measurement noise threshold $\delta
n >$ 0.5 cm$^{-3}$ (blue, 119,512 points total).}
\end{figure}

We compute the normalized, zero-lag cross-correlation $C(\delta n,
\delta B_{\parallel}) = \langle \delta n \delta B_{\parallel} \rangle/$\dnr\dbr,
which has a range from $-1$ to $1$.  As described qualitatively above,
we expect that $C(\delta n, \delta B_{\parallel})$ will be negative
(positive) for slow- (fast-) mode MHD fluctuations (see Figure 1).
Figure~2 shows the distribution of $C(\delta n, \delta
B_{\parallel})$, both for all of the data and for the restricted
dataset that exceeds the density noise threshold \dnr $ \ge$ 0.5~cm$^{-3}$.  While the entire dataset peaks below 0 (at \XCC $\simeq$
-0.5), the data with well resolved density amplitude levels peaks at
\XCC $\simeq$ -0.9.  Figure 3 shows the joint histogram of \XCC
vs.~ion plasma beta $\beta_i$.  The top panel shows the distribution
of points, with a histogram of $\beta_i$ overplotted.  The middle
panel is the joint histogram normalized to number of $\beta_i$ points
in each $\beta_i$ bin.  This shows clearly that \XCC is near $-1$ over
the entire interval, increasing slightly with $\beta_i$ to $\simeq -0.7$
at $\beta_i =10$.  In the bottom panel, the cumulative
distribution shows that fewer than 10\% of the intervals have \XCC$>0$.

\begin{figure}
\resizebox{3.65in}{!}{\includegraphics{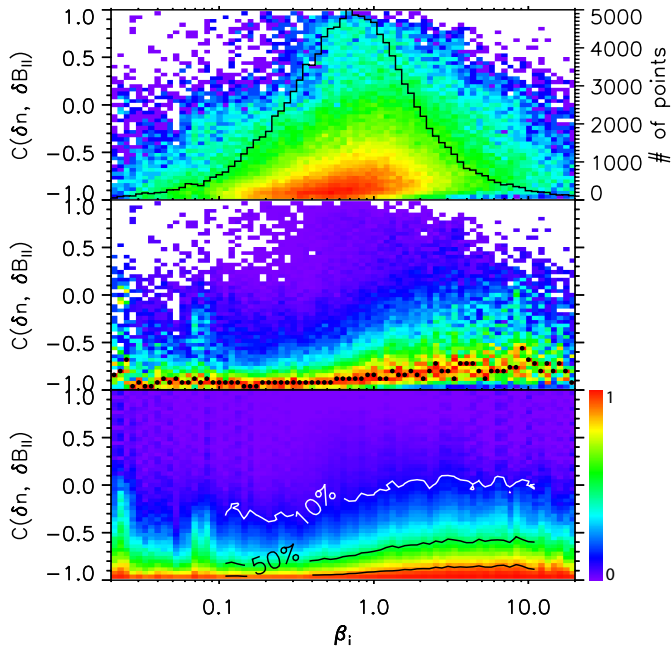}}
\caption{ \label{fig:xcc} (Upper) Total distribution of the  
$C(\delta n, \delta B_{\parallel})$ cross-correlation as a function of ion
plasma beta $\beta_i$.  The count in each $\beta_i$ bin is overplotted
(with the scale on the  right).  (Middle) Joint distribution of the
$C(\delta n, \delta B_{\parallel})$ cross-correlation normalized within each
$\beta_i$ bin.  Black dots are the peak values in each $\beta_i$ bin.
(Lower) Cumulative distribution of $C(\delta n, \delta B_{\parallel})$ with
contours at 90\%, 50\%, and 10\%.  These distributions consist of 119,512
independent data intervals.}
\end{figure}

\emph{Synthetic Data.}--- 
A cubic synthetic plasma volume spanning scales $3 \times 10^{-3} \le
k \rho_i \le 4.8 \times 10^{-2}$ is constructed using a $32^3$ grid.
A spectrum of linear waves, with 90\% of the energy in Alfven waves
and the remaining 10\% in a mixture of kinetic fast and slow waves,
consistent with the observed $k^{-5/3}$ one-dimensional energy
spectrum of the magnetic field fluctuations $|\delta \V{B}|$, is
created in the volume using the linear eigenfunctions for these modes
from the Vlasov-Maxwell linear dispersion relation
\cite{Quataert:1998,Howes:2006}; see Klein \emph{et al.}~2011
\cite{Klein:2011} for more details. A fully-ionized proton and
electron plasma is assumed, with isotropic Maxwellian velocity
distributions, a realistic mass ratio $m_i/m_e=1836$, equal ion and
electron temperatures $T_i=T_e$, and non-relativistic conditions
$v_{ti}/c = 10^{-4}$.  Taking the MHD limit $k\rho_i \ll 1$, under
these conditions the normalized linear Vlasov-Maxwell eigenfrequency
depends on only two parameters, $\omega/(k v_A) =
\overline{\omega}(\beta_i, \theta)$, the ion plasma beta $\beta_i$ and
the angle $\theta$ between the wavevector and the mean magnetic field 
\cite{Klein:2011}.  Once $\beta_i$ has been chosen, one
needs only to specify  the distribution of energy in wavevector
space. Compressible MHD turbulence simulations generate an isotropic
distribution of fast waves and critically balanced distributions of
\Alfven and slow waves \cite{Cho:2003}.  Therefore, we initialize the fast wave
energy isotropically, while the \Alfven and slow wave energy mimics a critically
balanced distribution by setting all modes with $k_\parallel >
k_0^{1/3} k_\perp ^{2/3}$ to zero, where $k_0$ corresponds to the
scale of the plasma volume.

Time series of density and parallel magnetic field fluctuations are
constructed by sampling the synthetic data at a probe moving through
the volume at an oblique angle with respect to the mean field (tests
have confirmed insensitivity to the choice of angle). We then compute
the cross-correlation $C(\delta n, \delta B_\parallel)$ as above.
Figure 4 shows $C(\delta n, \delta B_\parallel)$ for several values of
the ratio of fast wave energy to total compressible energy $F$ vs.~ion plasma
$\beta_i$.  Peak histogram values (and FWHM error bars) are also
shown.  The solar wind data are in striking agreement with the
synthetic data $F = 0.00$ curve.

\emph{Discussion.}--- Figure 4 shows that the
observed correlation is consistent with a statistically negligible
kinetic fast wave energy contribution for the large sample used in this
study. Note, however, that a very small fraction of the data intervals have a positive cross-correlation (see \figref{fig:histo}),
possibly indicating the presence of kinetic fast waves in these intervals.

\begin{figure}
\resizebox{3.4in}{!}{\includegraphics{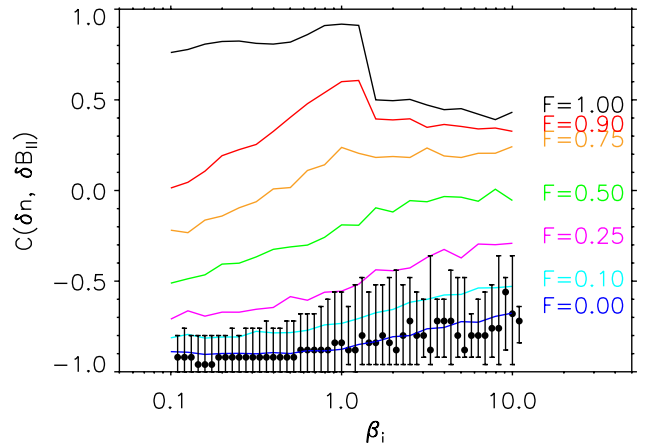}}
\caption{ \label{fig:money} Comparison of measured values of the 
$C(\delta n, \delta B_\parallel)$ cross-correlation (black dots with
FWHM error bars) to the synthetic data predictions for 
the ratio of kinetic fast wave to total compressible energy $F$.
Best agreement is with $F=0.00$, indicating that the compressible
component of solar wind turbulence is almost entirely in the kinetic
slow mode.}
\end{figure}

Previous analyses have generally dismissed the possibility of
kinetic slow waves because, in an isotropic Maxwellian plasma with
warm ions, the collisionless damping via free-streaming along the
magnetic field is strong \cite{Barnes:1966}. However, in the limit
$k_\perp \gg k_\parallel$ applicable to a critically balanced power
distribution, the damping rate of the slow waves is proportional to
the parallel component of the wavevector, $\gamma \propto k_\parallel$
\cite{Howes:2006}.  For exactly perpendicular wavevectors, the damping rate drops to
zero---this perpendicular limit of the slow wave corresponds to an
undamped, non-propagating pressure-balanced structure (PBS). In
compressible, strong MHD turbulence, it has been shown that the slow
modes are cascaded passively by the \Alfvenic turbulence
\cite{Maron:2001,Schekochihin:2009}, so the energy cascade rate is
related not to the slow wave frequency, but to the \Alfven wave
frequency. Therefore, the more nearly perpendicular slow waves
(possibly with $k_\parallel$ well below critical balance, $k_\parallel
\ll k_0^{1/3} k_\perp^{2/3}$) may be cascaded to smaller scales 
on the timescale of the \Alfvenic turbulence, while the collisionless
damping of these modes remains weak. 

We offer the following physical model of the compressible fluctuations
in solar wind turbulence.  At inertial range scales $k \rho_i \lesssim
0.1$, the density and parallel magnetic field fluctuations arise
mainly from the kinetic counterparts of the fast and/or slow MHD
waves. The measured \XCC cross-correlation at these scales
demonstrates that the compressible fluctuations are statistically
dominated by kinetic slow mode fluctuations, and the distribution of
power in wavevector space of these slow modes mimics the critically
balanced distribution expected of \Alfvenic fluctuations
\cite{Klein:2011}.  These kinetic slow wave fluctuations may be
cascaded as passive fluctuations to smaller scales by the \Alfvenic
turbulence
\cite{Maron:2001,Schekochihin:2009}, and so are predicted to exist
down to the scale of the ion Larmor radius, and perhaps to even
smaller scales. Thus, the evidence for PBSs over a range of timescales
from $10^3$~s to 10~s \cite{Yao:2011} is explained by the presence of a
distribution of kinetic slow waves that is undergoing a turbulent
cascade to smaller scales driven by the \Alfvenic turbulence.  In
addition, these kinetic slow modes suffer collisionless damping at a
rate $\gamma \propto k_\parallel$, meaning that the more perpendicular
the wavevector, the slower the damping rate.
 

\begin{figure}
\resizebox{3.35in}{!}{\includegraphics*[0.32in,4.75in][7.95in,9.7in]
{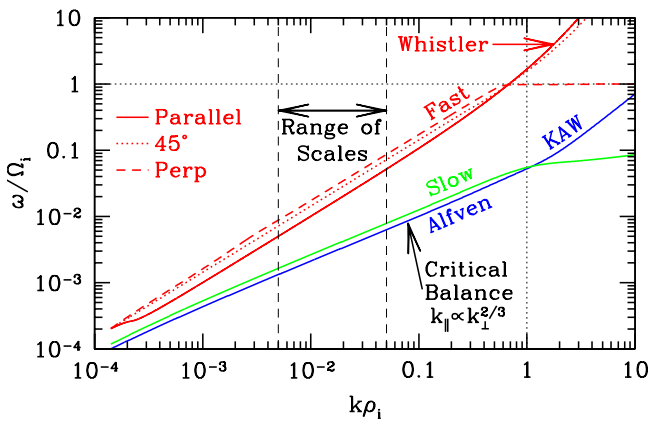}}
\caption{ \label{fig:om_krhoi}  Plot of frequency $\omega/\Omega_i$ 
vs.~wavenumber $k\rho_i$ for the collisionless versions of the fast
MHD (red), \Alfven (blue), and slow MHD (green) waves determined using
the Vlasov-Maxwell linear dispersion relation.  The range of scales
considered in this study is indicated.  Our measurements suggest that
the solar wind spectrum lies on the blue-green curves.}
\end{figure}

The lack of a statistically significant fast wave component in the
inertial range of solar wind turbulence has implications for
the cascade of energy to small scales.
Because the nonlinear energy transfer is believed to be dominated by
local interactions in wavenumber space, significant nonlinear energy
transfer occurs only between waves with similar linear frequencies.
\figref{fig:om_krhoi} shows the real linear frequencies $\omega$ of the 
collisionless counterparts of the MHD fast, Alfv\'en, and slow waves as
a function of $k \rho_i$.  For the isotropically distributed fast
waves (red), we plot the parallel (solid), $45^\circ$ (dotted), and
perpendicular (dashed) increase of the wavevector; slow (green) and
\Alfven (blue) waves follow the critically balanced 
path $k_\parallel =k_0^{1/3}k_\perp^{2/3}$, with the isotropic driving
scale $k_0 \rho_i = 10^{-4}$ for all cases. Since only the fast wave
turbulent cascade is expected to nonlinearly transfer energy to
whistler waves at $k \rho_i
\gtrsim 1$, our analysis suggests that there is little or no 
transfer of large scale turbulent energy through the inertial range
down to whistler waves at small scales.


Supported by NASA grants NNX10AC91G (Iowa) and NNX10AT09G (Berkeley)  and by 
NSF grants AGS-1054061 (Iowa) and AGS-0962726 (Berkeley).

%

%
\end{document}